\documentclass{appolb}
\usepackage{epsfig}

\newcommand{\be}{\begin{equation}}
\newcommand{\ee}{\end{equation}}
\newcommand{\ba}{\begin{eqnarray}}
\newcommand{\ea}{\end{eqnarray}}
\newcommand{\mean}[1]{\left\langle #1 \right\rangle}

\begin{document}

\title{Log-periodic oscillations in degree distributions of hierarchical scale-free networks}
\author{Krzysztof Suchecki and Janusz A. Ho{\l}yst\footnote{Corresponding author: jholyst@if.pw.edu.pl}
\address{Faculty of Physics and Center of Excellence for Complex Systems Research \\
Warsaw University of Technology\\
Koszykowa 75, PL--00-662 Warsaw, Poland}}
\date{\today}

\maketitle

\begin{abstract}
Hierarchical models of scale free networks are introduced where
numbers of nodes in  clusters of a given hierarchy are stochastic
variables.  Our  models show periodic oscillations of degree
distribution $P(k)$ in the log-log scale. Periods and amplitudes
of such oscillations depend on network parameters. Numerical
simulations are in a good agreement to analytical calculations.
\end{abstract}


\section{Introduction}
Recently there is  a large interest in scale-free networks that
seem to be good approximations for such systems as the Internet,
World Wide Web,  social or biological networks; for a review see
\cite{statmech}-\cite{book2}. A simple model that exhibits the
power law for degree distributions $P(k)$ observed in real complex
networks is the Barab\'{a}si-Albert model of preferential
attachment \cite{statmech}. The model however suffers from very
low values of the clustering coefficient $C$ \cite{FFH03} for
large networks as compared to observations of real systems
\cite{statmech}-\cite{book2}. To overcome this discrepancy a model
of hierarchical networks  has been introduced by Ravasz and
Barab\'{a}si (RB) where the clustering coefficient is much larger
\cite{Barabasi_hierarchy}. The RB network consists of
hierarchically connected clusters where numbers of nodes in every
cluster of a given hierarchy are the same.
The degree distribution $P(k)$ in this approach also exhibits power-law.
However, it is only a general trend. In fact, the degree distribution
consists of delta-peaks for only a few degree values, instead of continuous
distribution observed in real networks.
In this paper, we introduce a class of more general models, where number
of nodes in every cluster is a stochastic variabl, what seems to be more
justified for real network models. As result the peaks of $P(k)$ are blurred,
creating a network with wide range of possible $k$ values, but the log-periodic
behaviour of $P(k)$ is still clearly visible.\\
Let us remind that log-periodic oscillations are
characteristic features of systems where a discrete
self-similarity is present \cite{discrete_scaling} and the effect
can occur even without a preexisting hierarchy
\cite{discrete_scaling} in such various systems as  earthquakes
\cite{earth1,earth2} or financial markets
 where log-periodic oscillations were
observed as possible precursors for financial crashes
\cite{econo1,econo2,econo3}. Such oscillations were also found for
mean residence times at chaotic crisis where a collision of a
fractal attractor with a fractal or a nonfractal basin of another
attractor takes place \cite{kacperski,kacperski2} and for the
stochastic resonance in chaotic systems near a crisis point
\cite{krawiecki}.

\section{The Model}

\begin{figure}
\center{\epsfig{file=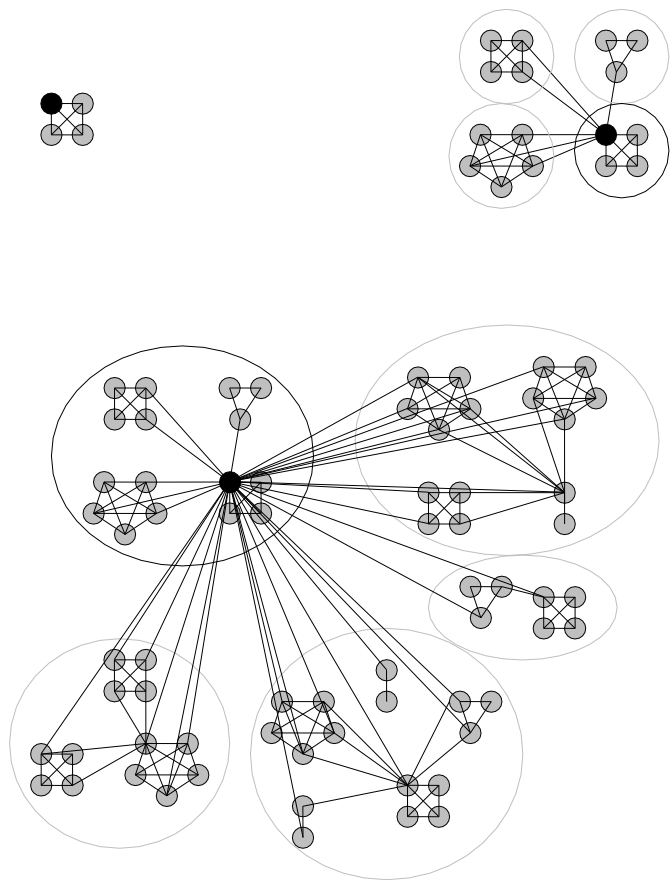,width=.7\textwidth,angle=-0}}
\caption{\label{netmodel} The first three steps of network
creation in the P1 model.}
\end{figure}

Our model possesses two parameters, a distribution $P_M(m)$, where
$m=1,2,3,\dots$ and a number $p \in (0,1]$. We start out from a
single cluster (a cluster of hierarchy $0$) of $m+1$ fully
connected nodes (Fig.\ref{netmodel}), where $m$ is a random number
from a distribution $P_M(m)$. One node in the cluster is its {\it
central} node. The central node of the cluster is a center of
hierarchy $0$. Next, we call our cluster the central one and
create a random number $m$ of similar clusters. Each is created in
the same way as the central cluster, but we pick a random number
$m$ for each one independently, therefore they may include
different numbers of nodes. Next, we connect a part $p$ of all
nodes in non-central clusters to the central node in the central
cluster. This node becomes the central node for the whole cluster
of hierarchy $1$ we have obtained so far. Similarly the central
node of our cluster is a center of hierarchy $1$. We repeat the
process, until we get a network of a desired hierarchy. This model
is referred to as P1 model. The model is generalization of the
stochastic model proposed by Barab\'{a}si and Ravasz
\cite{Barabasi_hierarchy}. If we take $P_M(m)=\delta(m,m_0)$,
where $m_0$ is constant, our model simplifies to BR model, with
number of nodes and degree distribution determined strictly by $p$
and $m_0$ values.

A variation of the model has been also studied. In each hierarchy
$d$ we connect not a fraction $p$ of nodes but a fraction $p^d$.
This model is referred to as the PD model.
\section{Degree Distribution of P1 and PD models}

\begin{figure}
\center{\epsfig{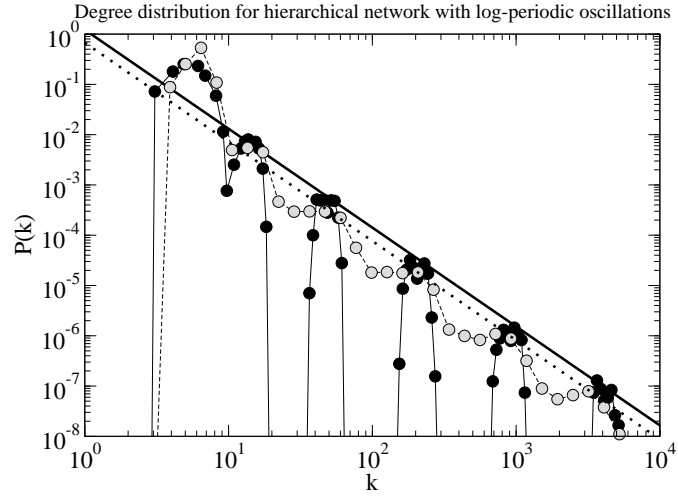}}
\caption{\label{blur} Degree distribution for two networks with
different $P_M(m)$. Filled circles are for $m=3$ or $4$ with equal
probability, gray circles are for uniform $m$ distribution between
$1$ and $5$. The straight lines show scaling of peak heights and
correspond to $\gamma=0.967$ for the first case (continuous line)
and $\gamma=0.973$ for the second (dotted line).}
\end{figure}

As previously noted, for $P_M(m)=\delta(m,m_0)$ we get a degree
distribution identical to that of BR model. It consists of
separate peaks, corresponding to degrees of central nodes of
following hierarchies. Central nodes of given hierarchy have a
fixed degree, dependent only on the network parameter $p$. At the
logarithmic scale the distance between neighboring peaks is
approximately constant and equals to $\log(m_0+1)$. The peaks
follow laws of discrete scaling \cite{discrete_scaling}. The
heights of peaks with degrees $k_i$ decrease as $k_i^{-\gamma}$,
and distances between consecutive peaks fulfill the relation
$k_{i+1}/k_i=\lambda$. The probability $P(k)$ between peaks equals
to zero what means that only nodes with peculiar degrees are
possible. But what happens when the number $m$ is not a fixed
value ?\\ Numerical simulations show that each peak blurs,
depending on the $P_M(m)$ distribution. If the blur is small, the
distribution consists of separate peaks, but they are not
delta-shaped. If the blur is large enough, the peaks overlap and a
continuous degree distribution is obtained. Figure \ref{blur}
shows degree distributions for both cases in P1 model. Both
display a discrete scaling, and have the same scaling exponent (up
to statistical fluctuations), independent on network parameters.\\
Similar behavior has been observed in the PD model, although
scaling exponent is parameters dependent.

\section{Mean $m$ value approach}
Numerical simulations have shown that when $m$ is not a constant
but a random number from a given distribution, peaks blur,
eventually overlapping and creating a continuous distribution.
However, regardless of the actual shape, the distribution still
consists of peaks. Each peak has an average degree $k$ and a mass
$n$ representing number of nodes that belong to this peak. All
nodes in a peak are centers of the same hierarchy. Using a {\it
mean value} of $m$, the distance between peaks and their relative
heights can be easily found. From these two values we directly get
the discrete scaling ratio $\lambda$ and the scaling exponent
$\gamma$. In the following calculations we neglect the degree
increase of nodes due to their connections to the central cluster,
as this effect increases the node degree at most by $d$, what is
insignificant for higher hierarchy centers.

\subsection{P1 model}
Let us denote an average degree of peak of hierarchy $d$ by $k_d$,
an average number of nodes in a cluster of hierarchy $d$ by $N_d$,
and an average number of centers of hierarchy $d$ in a network of
hierarchy $h$ by $n_d^h$. The network size $N_d$ increases
exponentially with hierarchy $d$ as $N_d=\mean{m+1}^{d+1}$.
Centers of hierarchy $0$ have a degree $k_0$ equal to $\mean{m}$
and it increases by $p \cdot \mean{m} \cdot N_{d-1}$ in each next
hierarchy $d$. We obtain \be k_d=\mean{m}+p\cdot \mean{m+1}\cdot
\left(\mean{m+1}^d-1\right) \ee

If $\mean{m+1}>1$ and $d>>1$, the above expression can be
simplified to \be \label{kd_form} k_d\approx p\cdot
\mean{m+1}^{d+1} \ee If the condition is not satisfied, distances
between peaks are not constant at logarithmic scale and the
network is not scale-free.

Since the discrete scaling ratio $\lambda$ simply equals
$k_{d+1}/k_d$ thus we get $\lambda \approx \mean{m+1}$.

The scaling exponent $\gamma$ can be found using the cumulative
degree distribution. Starting from \be \label{cumeq}
P(k)=\frac{\Delta P_{cum}}{\Delta k}=\frac{\Delta P_{cum}}{\Delta
d}\frac{\Delta d}{\Delta k} \ee where $d$ are consecutive
hierarchies, and using calculations presented in Appendix
\ref{app1} we get $P(k)\sim k^{-2}$ so the scaling exponent
$\gamma$ equals to $2$, regardless of $p$ and $P_M(m)$. Note that
this scaling is valid for peak masses $n_d^h$ only.

\subsection{PD Model}
The case of PD model is very similar to the P1 model. However,
since instead of a fraction $p$ we connect a fraction $p^d$ of
nodes from non-central clusters, the degree $k_d$ is
\be
\label{kd_pdform}
k_d=\mean{m}\frac{1-(p\cdot\mean{m+1})^{d+1}}{1-p\cdot\mean{m+1}}
\ee

When we assume that $p\cdot\mean{m+1}>1$ we can omit one in the
numerator and get the discrete scaling ratio
$\lambda=k_{d+1}/k_d\approx p\cdot\mean{m+1}$. Similarly to the P1
model, if it is not true, the network is not scale-free.

To find the scaling exponent, again we use cumulative degree
distribution and Eq.\ref{cumeq}. For the PD model we get
$\gamma=1+\frac{\ln\mean{m+1}}{\ln p\mean{m+1}}$. Note that since
$p\in(0,1]$ and $p\mean{m+1}>1$ for scale-free networks, the
scaling exponent is always greater than $2$.

\subsection{Numerical Data}

Numerical simulations have been performed for networks of
hierarchy $6$, with $p=0.5$ and various uniform distributions of
$m$, to find out if analytic predictions are correct.

\begin{figure}
\center{\epsfig{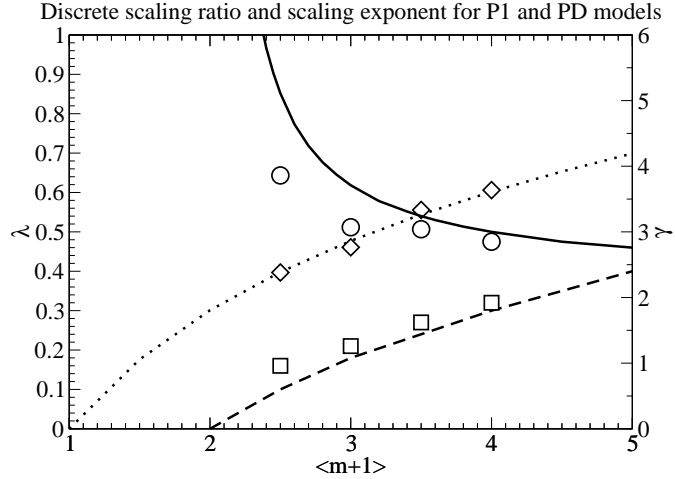}}
\caption{\label{lambda_gamma} The discrete scaling ratio $\lambda$
and the scaling exponent $\gamma$. Symbols are values obtained
from numeric simulations, lines are analytic predictions. Diamonds and
dotted line are for $\lambda$ in P1 model. Squares and dashed
line are for $\lambda$ in PD model. Circles and smooth line are
for $\gamma$ in PD model. In all cases, networks are of hierarchy
$6$, with parameter $p=0.5$.}
\end{figure}

Tables \ref{tab_p1} and \ref{tab_pd} contain obtained data. Figure
\ref{lambda_gamma} shows the comparison between prediction and
results.

As it can be seen the numerical data are in a good agreement with
our analytic predictions. The largest deviation is for low
$\mean{m}$ and for low $p$, where our approximations were poor.

\begin{table}
\center{\begin{tabular} {cccccc}
\hline $m$ & $\mean{m+1}$ & $\gamma_{analyt}$ &
$\gamma_{numer}$ & $\log \lambda_{analyt}$ & $\log
\lambda_{numer}$ \\
\hline
1 to 2 & 2.5 & 2 & 1.981 & 0.398 & 0.397 \\
1 to 3 & 3 & 2 & 1.978 & 0.477 & 0.461 \\
1 to 4 & 3.5 & 2 & 1.931 & 0.544 & 0.556 \\
1 to 5 & 4 & 2 & 1.973 & 0.602 & 0.606 \\
\hline
\end{tabular}}
\caption{\label{tab_p1} Distribution of $m$ numbers, analytic and
numerical scaling exponents $\gamma$ and the logarithm of discrete
scaling ratio $\lambda$ for the model P1. Data obtained from
averaging over $30$ networks.}
\end{table}

\begin{table}
\center{\begin{tabular} {cccccc}
\hline
$m$ & $\mean{m+1}$ & $\gamma_{analyt}$ & $\gamma_{numer}$ & $\log \lambda_{analyt}$ & $\log \lambda_{numer}$\\
\hline
1 to 2 & 2.5 & 5.106 & 3.858 & 0.097 & $\sim$0.16\\
1 to 3 & 3 & 3.710 & 3.067 & 0.176 & 0.208\\
1 to 4 & 3.5 & 3.239 & 3.038 & 0.243 & 0.271\\
1 to 5 & 4 & 3.000 & 2.846 & 0.301 & 0.32\\
\hline
\end{tabular}}
\caption{\label{tab_pd} Distribution of $m$ numbers, analytic and
numerical scaling exponents $\gamma$ and the logarithm of discrete
scaling ratio $\lambda$ for the model PD. Data obtained from
averaging over $2000$ networks. In the first row, the exact
$\lambda$ value was impossible to obtain, due to very weak
periodic behavior.}
\end{table}

\section{Exact degree distribution}
Up to now, all calculations have been performed using only the
average $m$ value, treating the degree distribution as series of
peaks. We have been concentrating on relations between peak's
masses and distances, while ignoring their shape. Here we find a
shape of the degree distribution for the P1 model.\\ Let $P_M(m)$
be a distribution of $m$, where $m$ is a number of noncentral
clusters in each hierarchy. Let $\widetilde{P}_d(N)$ be a
distribution of the network sizes $N$ for hierarchy $d$. $P_d(k)$
is a degree distribution for a network of hierarchy $d$,
$P_d^c(k)$ is a degree distribution for the central node of
hierarchy $d$.

The number of nodes in the network can be found as follows.
Network of hierarchy $d=0$ has $m+1$ nodes what means
$\widetilde{P}_0(N)=P_M(N-1)$. The size of each next hierarchy
$d+1$ is a sum of $m+1$ independent values, which are sizes of
networks of hierarchy $d$. \be \label{PdN}
\widetilde{P}_{d+1}(N)=\sum_m
P_M(m)\sum_{n_1,n_2,\dots,n_m}\widetilde{P}_d(n_1)\widetilde{P}_d(n_2)\dots\widetilde{P}_d(N-n_m-\dots-n_1)
\ee This recursive formula describes the probability distribution
for the network size $N$.

A network of hierarchy $d=0$ has degree distribution
$P_0(k)=P_M(k)$. This distribution describes both regular nodes
and a center of hierarchy $0$, which have the same degree values.
In each next hierarchy $d+1$ the degree distribution for all nodes
of hierarchy $d$ or less is the same, since we omit the degree
increase due to connections to the central node of higher
hierarchy. Now, we multiply the distribution by $(m+1)$ and add
the degree distribution $P_{d+1}^c(k)$ for the central node of the
network. This way we obtain an unnormalized degree distribution
for the whole network of hierarchy $d+1$. \be
P_{d+1}(k)=\sum_m\left[(m+1)P_M(m)P_d(k)+P_{d+1}^c(k)\right] \ee
The center is roughly connected to fraction $p$ of all nodes in
the network, what means it possesses the degree $p\cdot N$. This
yields the distribution of its degree equal to
$P_1^c(k)=\widetilde{P}_1(k/p)$. As result we obtain \be
\label{Pdk}
P_{d+1}(k)=\sum_m\left[(m+1)P_M(m)P_d(k)+\widetilde{P}_{d+1}(k/p)\right]
\ee

This recursive formula describes the unnormalized degree
distributions for networks of consecutive hierarchies $d$, with
the exception of $d=1$. Since $P_0(k)$ describes not only centers
of hierarchy $0$ but both regular nodes and centers, we must
account for that. We do so by multiplying $P_0(k)$ in the formula
by the average basic cluster size $\mean{m+1}$.

\be \label{P1k}
P_1(k)=\sum_m\left[\mean{m+1}(m+1)P_M(m)P_0(k)+\widetilde{P}_1(k/p)\right]
\ee

In the above calculations, like in the calculations using average
$m$, we omitted the degree increase due to connections to the
central node of higher hierarchy. This is insignificant for higher
hierarchy centers, as the increase is at most $d$, while the
center degree increases exponentially with $d$. We have used the
formula $P_d^c(k)=\widetilde{P}_d(k/p)$ in the above calculations.
In reality degree distributions are discrete, with natural $k$
values. Depending on how we round the number of connections to the
central node, we should interpret the above formula accordingly.

We rounded the number of connections $p\cdot N$ down, what gives
the following formula for interpreting the probability with
fractional argument \be \label{Pfract} P_d^c(k)=\sum_{l\geq
k/p,l<(k+1)/p}\widetilde{P}_d(l) \ee It means that the probability
of getting the center of degree $k$ equals to the sum of
probabilities for $N$, that lead to this $k$. Along with \be
\label{P0N} \widetilde{P}_0(N)=P_M(N-1) \ee and \be \label{P0k}
P_0(k)=P_M(k) \ee Eqs.\ref{PdN} and \ref{Pdk}-\ref{Pfract} allow
to find numerically an exact but unnormalized degree distribution
for the P1 model.

\begin{figure}
\center{\epsfig{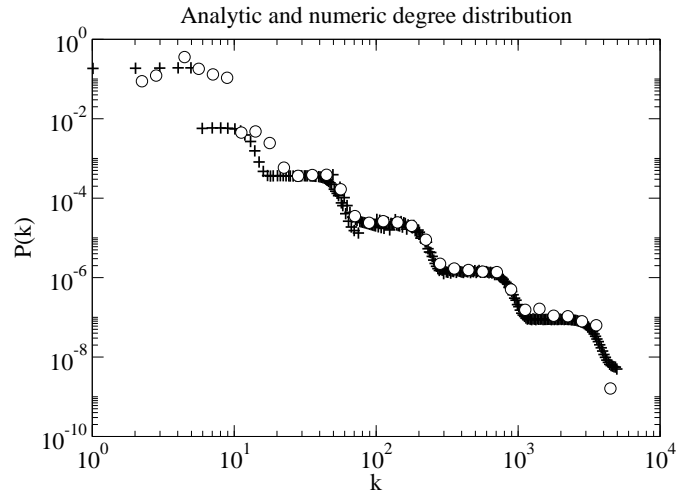}}
\caption{\label{exact15} Degree distribution for the network with
uniform $m$ distribution from $1$ to $5$, and hierarchy $d=5$. The
graph shows analytic (crosses) and numeric data (circles).}
\end{figure}

\begin{figure}
\center{\epsfig{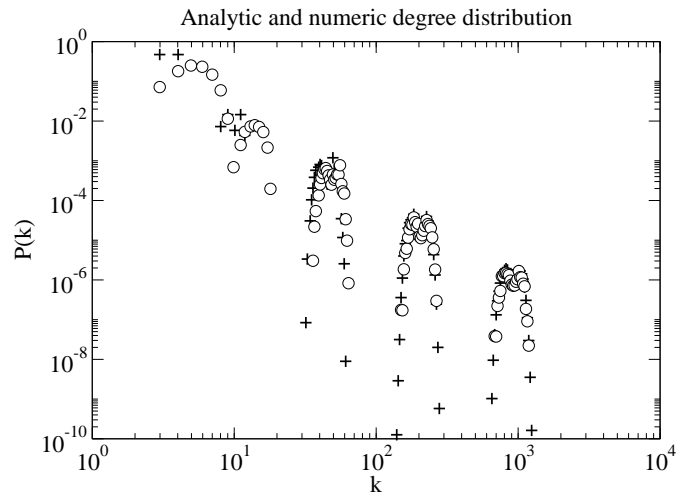}}
\caption{\label{exact34} The degree distribution for network with
the uniform $m$ distribution ($3$ or $4$), and the hierarchy
$d=4$. The graph shows analytic (crosses) and numeric data
(circles).}
\end{figure}

Comparing these formulas with numerical data one can see that our
calculations are correct for higher degrees, where approximations
we used are accurate (Fig.\ref{exact34}, Fig.\ref{exact15}).

\begin{figure}
\vskip 0.5cm
\center{\epsfig{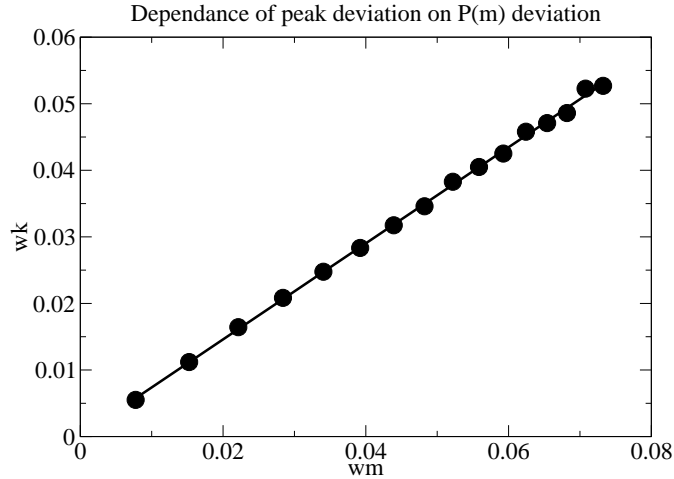}}
\caption{\label{odchylenia} Relation between the peak width $wk$
(a standard deviation of peaks in $P(\ln k)$) and the distribution
width $wm$ (a standard deviations of $P(\ln m)$). The slope of the
line is $\alpha=0.722$.}
\end{figure}

Using degree distributions obtained with our formulas
(Eq.\ref{PdN},Eqs.\ref{Pdk}-\ref{P0k}), a relation between the
distribution of $m$ and a peak shape has been found. We have
studied various uniform distributions of $m$ and have found linear
relation between the standard deviation of the distribution $P(\ln
m)$ and the standard deviation of peaks in the $P(\ln k)$
distribution (Fig.\ref{odchylenia}).

Peak deviations in the distribution P(k) are calculated at the
logarithmic scale of $k$ \be \sigma=\frac{\sum(\ln k-\mean{\ln
k})^2\cdot P(\ln k)}{\sum P(\ln k)}\ee Similar formula has been
used to calculate the deviation of $P(m)$. Peak deviations have
been calculated for the peak of the highest hierarchy. In the case
of overlapping peaks the minimums of $P(k)$ have been considered
borders of peak. The approximation is quite accurate, as $P(k)$
decays fast when we go away from the peak average $k$ value.

\section{Discussion}

The question occurs, whether our model corresponds to real network systems. It
is obvious that many real networks posess a hierarchical structure but of course
a detailed mechanism responsible for its emergence is unknown.
According to our knowledge, log-periodic oscillations around the power law in degree
distributions were never directly reported in the studies of real networks
or corresponding models. One can suspect however, that in many cases such oscillations
were visible and could be overlooked if the binning
or data averaging had been performed. Small amplitude oscillations can be also easily confused with
random fluctuations. The situation resembles oscillations around the scaling law in chaotic crises, where the periodic
part is also often omitted as fluctuations \cite{kacperski,kacperski2}.\\
A clear example of log-periodic oscillations for real networks can be seen in
the study of liability connections between Austrian banks \cite{banki}. As the authors stress
\cite{banki} a significant part of studied banking sector posesses a strong hierarchical
structure, what can be easily detected looking at a corresponding connection graph.
Two periods of oscillations can be identified at out-degree distribution
describing the number of liabilities to other Austrian banks (regardless of liability size) \cite{banki}.
The period of the oscillations is approximately $\lambda=k_{i+1}/k_i\approx 3$.
According to our theory, they are a result of the network's hierarchical structure.\\
We have found also far less visible oscillations in the studies of computer directory trees
\cite{katalogi} and World Trade Web \cite{worldtrade}, where the hierarchical structure can
be identified at a corresponding connection graph \cite{katalogi} or in dependence of a clustering coefficient on
a node degree \cite{worldtrade}.
Sizes of such oscillations are however at a fluctuations level. The other possible example can be found
in the paper \cite{japonczyk}, that
presents a non-monotenous behavior of degree distribution of P(k) for a shareholding network
in Japan. Here a single wave around the power law can be observed, where $\lambda=k_{i+1}/k_i\approx 10$.

\section{Conclusions}

In conclusion we have shown that hierarchical networks models
display log-periodic oscillations in the degree distribution when
the number of clusters forming the self-similar hierarchy is a
stochastic variable. The period and the amplitude of these
oscillations reflect the hierarchical structure of the network.
We also point out examples of real networks that display such
features.
It follows that observations of log-periodic oscillations in degree
distributions of real networks can give hints towards the
existence of hidden hierarchical structures in such systems.

\section{Acknowledgement}
This work has been partially supported by by a special Grant {\it
Dynamics of Complex Systems} of the Warsaw University of
Technology and by the EU Grant {\it Measuring and Modelling Complex Networks Across Domains} (MMCOMNET).

\appendix

\section{Mean $m$ value calculations} \label{app1}

This Appendix contains exact calculations regarding the discrete
scaling ratio $\lambda$ and the scaling exponent $\gamma$ for the
mean $m$ approach. The number $k_d$ is an average degree of peak
of hierarchy $d$, $N_d$ is an average number of nodes in network
of hierarchy $d$, $n_d^h$ is an average number of centers of
hierarchy $d$ in a network of hierarchy $h$. The number
$\mean{m+1}$ is a mean number of clusters in each hierarchy.

The Eq.\ref{kd_form} can be obtained as
\begin{eqnarray}
 \label{kd_full}
 k_d=k_{d-1}+\mean{m} \cdot N_{d-1} \cdot p=\mean{m}+p\cdot
 \mean{m} \cdot \sum ^{i=1}_{d} \mean{m+1}^i=\\
 =\mean{m}+p\cdot \mean{m+1}\cdot \left(\mean{m+1}^d-1\right) \nonumber
\end{eqnarray}

The scaling exponent $\gamma$ for model {\it P1} has been obtained
using the cumulative degree distribution \be \label{cumeq2}
P(k)=\frac{\Delta P_{cum}}{\Delta k}=\frac{\Delta P_{cum}}{\Delta
d}\frac{\Delta d}{\Delta k} \ee First we find an expression for
$n_d^h$. There is only one center of hierarchy $d=h$ in the
network of such a hierarchy. Each time the network hierarchy $h$
increases, the number of centers of hierarchy $d$ increases
$\mean{m+1}$ times. The exception is the first step, where one
node becomes center of the higher hierarchy. Because of that
$n_d^h$ increases only by the factor of $\mean{m}$ for that step.
We obtain $n_d^h=\mean{m}\cdot\mean{m+1}^{h-d-1}$ except for
$n_h^h=1$.

Now we calculate expressions in Eq.\ref{cumeq2}. Each next peak
is smaller $\frac{\Delta P_{cum}}{\Delta d}=n_d^h\sim\mean{m+1}^{-d}$ while its average
degree $k_d$ increases $\frac{\Delta k}{\Delta d}\sim\mean{m+1}^d$ In such a way we obtain \be \label{pk_full}
P(k)\sim \mean{m+1}^{-d}\cdot\mean{m+1}^{-d}=\mean{m+1}^{-2d}\ee

From Eq.\ref{kd_full} we get $d\approx\log_{\mean{m+1}}\frac{k}{p}-1=\frac{\ln k - \ln p}{\ln
\mean{m+1}}-1$.
By putting so calculated $d$ into Eq.\ref{pk_full} we get
\begin{eqnarray}
P(k)\sim\mean{m+1}^{-2d}=\exp\left(-2d\ln\mean{m+1}\right)= \\
\exp\left(-2\frac{\ln k - \ln
p}{\ln\mean{m+1}}\ln\mean{m+1}+2\ln\mean{m+1}\right)\sim\exp\left(-2\ln
k\right)=k^{-2} \nonumber
\end{eqnarray}

For the {\it PD} model the Eq.\ref{kd_pdform} can be obtained in
the following way
\begin{eqnarray}
\label{kd_pdfull}
k_d=k_{d-1}+\mean{m}\cdot N_{d-1}\cdot p^d=\\
=\mean{m}\cdot\sum_{i=0}^d
p^i\cdot\mean{m+1}^i=\mean{m}\frac{1-(p\cdot\mean{m+1})^{d+1}}{1-p\cdot\mean{m+1}}
\nonumber
\end{eqnarray}

The scaling exponent has been calculated in a similar way to the
case of model P1. The slope $\Delta P_{cum}/\Delta d$ is the same
as in the previous case but $\Delta k/\Delta
d\sim(p\cdot\mean{m+1})^d$ thus $P(k)\sim
p^{-d}\cdot\mean{m+1}^{-2d}$.

Expressing $d$ by $k$ we get
$d=\log_{p\mean{m+1}}\frac{(p\cdot\mean{m+1}-1)\cdot k}{\mean{m}}-1$
and putting that into Eq.\ref{cumeq2} we get $P(k)\sim
k^{-(1+\frac{\ln\mean{m+1}}{\ln p\mean{m+1}})}$ which yields
exponent $\gamma=1+\frac{\ln\mean{m+1}}{\ln p\mean{m+1}}$ for PD model.

\end{document}